\documentclass{article}
\usepackage{dcase2020_techrep,amsmath,graphicx,url,times,booktabs,tabularx,bm}

\usepackage{amsfonts,cite}
 
\DeclareMathOperator{\betadist}{Beta}

\title{The NTT DCASE2020 Challenge Task 6 system: Automated Audio Captioning with Keywords and Sentence Length Estimation}
\name{Yuma Koizumi, Daiki Takeuchi, Yasunori Ohishi, Noboru Harada, and Kunio Kashino}
\address{
NTT Corporation, Japan
}

\begin{document}

\ninept
\maketitle

\begin{sloppy}

\begin{abstract}
This technical report describes the system participating to the Detection and Classification of Acoustic Scenes and Events (DCASE) 2020 Challenge, Task 6: automated audio captioning. Our submission focuses on solving two indeterminacy problems in automated audio captioning: word selection indeterminacy and sentence length indeterminacy. We simultaneously solve the main caption generation and sub indeterminacy problems by estimating keywords and sentence length through multi-task learning. We tested a simplified model of our submission using the development-testing dataset. Our model achieved $20.7$ SPIDEr score where that of the baseline system was $5.4$.
\end{abstract}

\begin{keywords}
Audio captioning, sequence-to-sequence model, keyword estimation, acoustic event/scene estimation.
\end{keywords}

\section{Introduction}
\label{sec:intro}

This technical report describes the system participating to the Detection and Classification of Acoustic Scenes and Events (DCASE) 2020 Challenge, Task 6: automated audio captioning \cite{task}.
Automated audio captioning (AAC) is an intermodal translation task when translating an input audio into its description using natural language \cite{ac1,ac2,ac3,audiocaps,clotho}. In contrast to automatic speech recognition (ASR), which converts a speech to a text, AAC converts environmental sounds to a text. This task potentially raises the level of automatic understanding of sound environment from merely tagging events \cite{aed,aed2} (e.g.\,alarm), scenes \cite{asc} (e.g.\,kitchen) and condition \cite{asd} (e.g.\,normal/anomaly) to higher contextual information, for example, ``{\it a digital alarm in the kitchen has gone off three times}.''

Our submission focuses on solving the indeterminacy problems in AAC which were tackled in our previous studies \cite{ac2,tracke}. This indeterminacy can be broadly divided into the indeterminacy in (i) word selection \cite{tracke} and (ii) sentence length \cite{ac2}. The first problem is caused by that one acoustic event/scene can be described with several words, such as \{{\it car, automobile, vehicle, wheels}\} and \{{\it road, roadway, intersection, street}\} \cite{tracke}. The second one is caused by that a sound can be explained in either short or long sentences, such as ``noisy car sounds,'' or ``a lot of cars are driving on the roadway and there are very loud engine noises'' \cite{tracke}. Such indeterminacy leads to a combinatorial explosion of possible answers, making it almost impossible to estimate the ground-truth and difficulty in training an AAC system.

Our strategy for solving these problems is to simultaneously estimate keywords and sentence length through multi-task learning framework. Figure \ref{fig:sysov} shows the overview of our system. The pre-processing stage involves rule-based keywords and sentence length extraction from the caption and metadata. The captioning DNN has keyword estimation and a sentence length estimation branches, and estimates the ground-truth caption by integrating these results.

\begin{figure}[t]
  \centering
\includegraphics[width=85mm,clip]{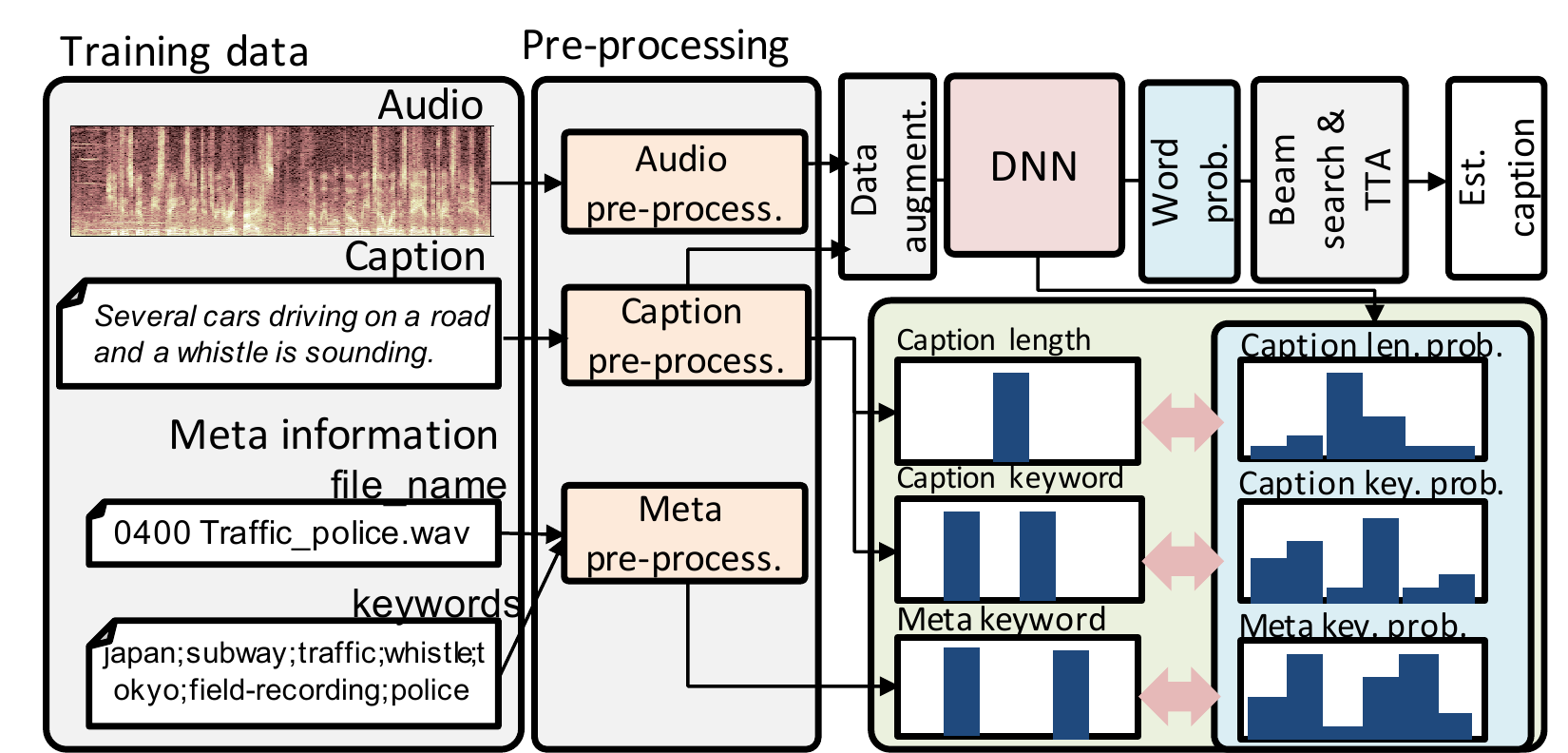} 
  \vspace{-15pt}
  \caption{System overview.}
  \label{fig:sysov}
  \vspace{-10pt}
\end{figure}

\section{System description}

This section describes the detail of our system. Since this paper is a technical report, we focus on describing the detailed implementation of the system. Effectiveness of each modules will be discussed in the workshop paper through ablation studies.

%%%%%%%%%%%%%%%%%%%%%%%%%%%%%%%%%%%%%%%%%%%%%%%%%%%%%%%%%%%
\subsection{Pre-processing}

%%%%%%%%%%
{\bf Audio pre-processing:}
As acoustic feature, we used three log-mel-spectrograms calculated from the time-domain input audio $\bm{x}$.
The first one was the log-mel-spectrogram of the input audio $\bm{S} \in \mathbb{R}^{F \times T_s}$,
where $F$ and $T_s$ are the number of mel-filterbanks and time-frames.
The second and third ones were that of the harmonic-percussive source separation (HPSS) outputs, $\bm{H} \in \mathbb{R}^{F \times T_s}$ and $\bm{P} \in \mathbb{R}^{F \times T_s}$.
These three spectrograms were concatenated on the channel dimension $\bm{X} \in \mathbb{R}^{3 \times F \times T_s}$.

The hyper-parameters of the audio pre-processing are as follows.
All audio samples were down-sampled at 22.05 kHz.
The window- and hop-size of short-time Fourier transform (STFT) were 4096 and 2048 points, respectively.
The number of mel-filterbank was $F=64$.
The hyper-parameters of the HPSS were default one of \url{librosa.decompose.hpss} \cite{hpss}.

%%%%%%%%%%
\vspace{3pt}
\noindent
{\bf Caption pre-processing:}
All captions were tokenized using the word tokenizer of the natural language toolkit (NLTK) \cite{nltk} while removing punctuation. All tokens in the development dataset were then counted, and words that appeared more than five times were appended in the word vocabulary. The vocabulary size was $C^{\mbox{\scriptsize cap}} = 2144$, which includes BOS, EOS, PAD, and UNK tokens.
In addition, the sentence length $L$ of each caption was counted.
Also, caption keywords $\bm{k}^{\mbox{\scriptsize cap}} = \{ k_i^{\mbox{\scriptsize cap}} \}_{i=1}^{K_c}$ was extracted using the keyword vocabulary which is discussed below.

%%%%%%%%%%
\vspace{3pt}
\noindent
{\bf Meta pre-processing:} 
Meta keywords were extracted from the \url{file_name} and \url{keyword} provided in the metadata csv file, using a keyword vocabulary which was manually created beforehand.
The procedure of creating the keyword vocabulary is as follows. First, \url{file_name} and \url{keyword} were split at places at space and punctuation. Next, words that seem to be nouns, verbs, adjectives, and adverbs were converted to its lemma. Finally, all lemmas were counted, and lemmas that appeared more than ten times were appended in the keyword vocabulary, which is a hash table that maps the original word to its lemma.
The vocabulary size was $C^{\mbox{\scriptsize key}} = 421$.
The keyword vocabulary was used to extract meta keyword $\bm{m} = \{ m_k \}_{k=1}^{K_m}$ and caption keyword $\bm{c}$.
Note that the procedure for creating the keyword vocabulary can be automated by using the part-of-speech (POS)--tagger and the WordNet Lemmatizer of the NLTK, however, we did this manually because their use is prohibited in this task.

%%%%%%%%%%%%%%%%%%%%%%%%%%%%%%%%%%%%%%%%%%%%%%%%%%%%%%%%%%%
\subsection{Data augmentation}

%%%%%%%%%%
\vspace{3pt}
\noindent
{\bf TF-IDF-based sample selection and data augmentation:} Since the target metrics of this challenge is SPIDEr, we need to accurately predict captions which include low frequent words and topics.  
To deal with word and topic bias in the training dataset, we adopted two tricks for training sample selection based on inverse document frequency (IDF), and one trick for data augmentation based on term frequency (TF)--IDF \cite{uda}.

The first trick is for selecting an audio sample $\bm{x}$ from the training dataset.
First, we concatenated the five ground-truth captions corresponding to each $\bm{x}$ in the training dataset, and used as a ``sentence''.
Then, we calculated IDFs for all words in all sentences, and calculated the average IDF of each sentence.
Finally, each average IDF was normalized by the sum of the average IDF. We regarded the normalized IDF as the parameter of the Categorical distribution, and selected $\bm{x}$ based on this probability.

The second trick is for selecting a ground-truth caption $\bm{w}$ from five captions corresponding to the selected $\bm{x}$.
The basic strategy was the same as the first trick.
First, we calculated IDFs of all words in the five captions. Here, note that the document was the five captions in contrast to the first trick.
Then, we calculated the normalized IDF and used as the parameter of Categorical distribution, and selected the target caption $\bm{w}$ based on this probability.

Finally, we adopted the third trick which is the TF-IDF based word replacement \cite{uda} to augment text data.

%%%%%%%%%%
\vspace{3pt}
\noindent
{\bf Random data cropping:} 
To train our captioning DNN using mini-batches, we adjusted the input length of audio sequence and text sequence using random cropping and padding.
We set the input length of audio to 20 seconds ($T=216$), and the number of words is $N=20$.
Thus, the inputs of the captioning DNN were $\bm{X} \in \mathbb{R}^{3 \times F \times T}$ and $\bm{w} = (w_1,...,w_N)$.
For $\bm{X}$ whose $T_s$ was greater than $T$, a random crop was performed so that the time-length was $T$, and shorter ones were applied zero-padding.
Similarly, if the sentence length was greater than $N$, words after the $N$-th word were cropped, and PAD tokens were added for shorter ones.

%%%%%%%%%%
\vspace{3pt}
\noindent
{\bf Mix-up:} 
After adjusting input length $T$ and $N$, we used the mix-up data augmentation.
First, we drew a mixing parameter $\beta$ from a beta distribution as
$
\beta \sim \betadist( 0.4, 0.4 )
$
where $\sim$ is sample drawing from the right-hand distribution.
Then, two audio samples were mixed by multiplying $\beta$ and $(1-\beta)$, respectively.
Since text inputs a set of class labels, direct mixing of $\bm{w}$ is not suitable. Thus, we mixed the embedded word tokens by multiplying the mixing parameters.

%%%%%%%%%%%%%%%%%%%%%%%%%%%%%%%%%%%%%%%%%%%%%%%%%%%%%%%%%%%
\subsection{Model description}

\begin{figure}[t]
  \centering
\includegraphics[width=85mm,clip]{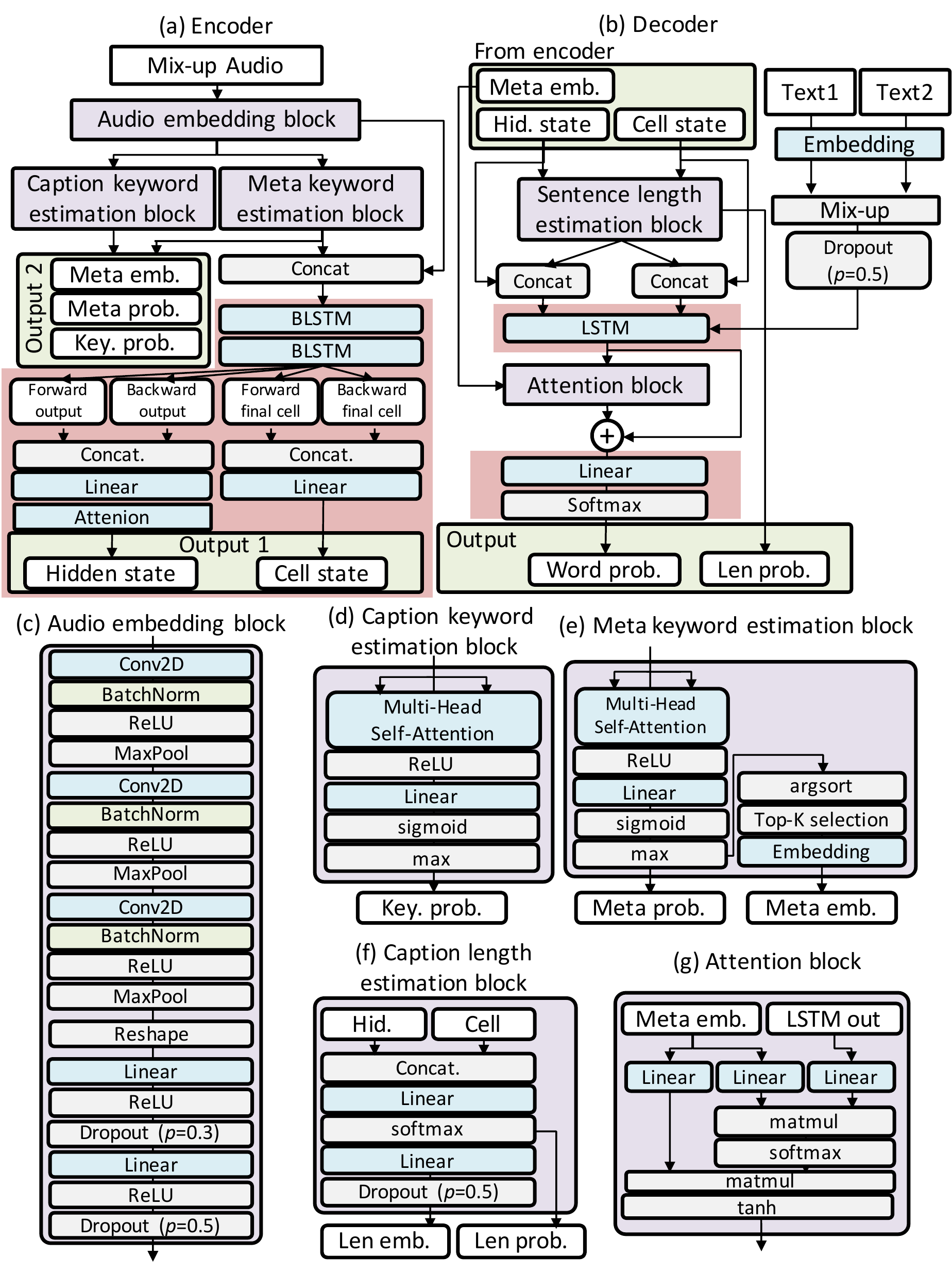} 
  \vspace{-15pt}
  \caption{Network architecture of captioning DNN.}
  \label{fig:modelov}
  \vspace{-10pt}
\end{figure}

Figure \ref{fig:modelov} shows the network architecture of the captioning DNN.
The pink area in Fig.\,\ref{fig:modelov} is a basic sequence-to-sequence (Sec2Sec)--based captioning model \cite{seq2seq1} using bidirectional long short-term memory (BLSTM)--LSTM. 
The encoder BLSTMs outputted the initial hidden state $\bm{h} \in \mathbb{R}^{D}$ and cell states $\bm{c} \in \mathbb{R}^{D}$ of the decoder LSTM, where $D=120$ was the hidden dimension of the whole network.
Then, the decoder LSTM estimated posterior probability of $n$-th word given the audio signal $\bm{x}$ and 1st to $(n-1)$-th words $p( w_n | \bm{x}, w_{1,..., n-1} )$ by using embedded word tokens.
In our submission, to solve the indeterminacy problems in AAC, we additionally used sub-blocks for keyword and sentence length estimation.
The following describes these sub-locks in detail.

%%%%%%%%%%
\vspace{3pt}
\noindent
{\bf Audio embedding block $\mathcal{A}$:} The input audio $\bm{X}$ was first passed to this block.
This block embeded $\bm{X}$ into a feature space as
$ \bm{A} = \mathcal{A}( \bm{X} ) \in \mathbb{R}^{D \times Ta}.
$
As shown in Fig.\,\ref{fig:modelov} (c), this block consisted of three convolutional neural network (CNN)--blocks and two fully connected (FC)--blocks.
The kernel size, stride, padding, and number of output channels of CNN were 3, 1, 1, and 64 for all CNN layers, respectively.
The kernel size and stride of the 2D max-pooling were 2 and 2, respectively.
Then, the output of CNN-blocks $64 \times F_a \times T_a $ was reshaped in to 
$ 64F_a \times T_a$, where $F_a = \frac{F}{2^3} = 8$ and $T_a = \frac{T}{2^3} = 27$, respectively.
The reshaped output was passed to the first FC layer which converts $\mathbb{R}^{ 64F_a \times T_a }$ to $\mathbb{R}^{ D \times T_a }$.
Finally, the second FC-layer outputted $\bm{A} \in \mathbb{R}^{D \times Ta}$.

%%%%%%%%%%
\vspace{3pt}
\noindent
{\bf Caption keyword estimation block $\mathcal{C}$:}
This block estimated
caption keyword probabilities of each keyword
$\bm{p}^{\mbox{\scriptsize cap}} \in [0,1]^{ C^{\mbox{\tiny key}}}$ from $\bm{A}$ as
$
\bm{p}^{\mbox{\scriptsize cap}} 
= \mathcal{C}(  \bm{A} ).
$
We expected that this block guides the audio embedding block so that its output includes information of the keywords of the ground-truth caption.
As shown in Fig.\,\ref{fig:modelov} (d), this block consisted of a muti-head self-attention (MHSA) layer and a FC layer.
The number of heads of MHSA was 4.
The output shape of MHSA and FC layer were $D \times T_a$ and $C^{\mbox{\scriptsize key}} \times T_a$, respectively.
Since the caption keyword has no time-labels, we aggregated the output by taking a maximum value in the time direction and outputted caption keyword probabilities of each keyword $\bm{p}^{\mbox{\scriptsize cap}}$.

%%%%%%%%%%
\vspace{3pt}
\noindent
{\bf Meta keyword estimation block $\mathcal{M}$:} 
This block estimated meta keyword probabilities $\bm{p}^{\mbox{\scriptsize meta}} \in [0,1]^{ C^{\mbox{\tiny key}}}$ and its embedding $\bm{M} \in \mathbb{R}^{ D \times K_m }$ from $\bm{A}$ as
$\{
\bm{p}^{\mbox{\scriptsize meta}}, \bm{M} \}
= \mathcal{M}(  \bm{A} ).
$
As shown in Fig.\,\ref{fig:modelov} (e), the base architecture was the same as the caption keyword estimation block, which consisted of MHSA and FC layers.
The base architecture outputted meta keyword probabilities of each keyword $\bm{p}^{\mbox{\scriptsize meta}}$.
To embed the estimated meta keywords into the feature space, first, we used the \url{argsort} function which returns the index on which $\bm{p}^{\mbox{\scriptsize meta}}$ sorts in descending order.
Then the top $K_m =15$ indices were selected as the estimated meta keyword $\{ \hat{m}_k \in \mathbb{N} \}_{k=1}^{K_m}$.
Finally, these indexes were passed to the embedding layer to obtain the estimated meta keyword embedding $\bm{M}$.
After this block, $\bm{A}$ and $\bm{M}$ were concatenated as $(\bm{M},\bm{A},\bm{M})$, and it was passed to BLSTMs.

%%%%%%%%%%
\vspace{3pt}
\noindent
{\bf Sentence length estimation block $\mathcal{L}$:} 
This block estimated the sentence length probability $\bm{p}^{\mbox{\scriptsize len}} \in [0,1]^{ L^{\mbox{\tiny max}} }$ and its embedding $\bm{l} \in \mathbb{R}^{D_l}$ as
$\{ \bm{p}^{\mbox{\scriptsize len}}, \bm{l} \} = \mathcal{L}( \bm{h}, \bm{c} ),
$ where 
$L^{\mbox{\tiny max}} = 20$ is the maximum sentence length that we assumed.
First, $\bm{h}$ and $\bm{c}$ were concatenated, 
and the first FC layer estemited $\bm{p}^{\mbox{\scriptsize len}}$ from the concatenated feature.
Finally, $\bm{p}^{\mbox{\scriptsize len}}$ was passed to the second FC layer, and outputted $\bm{l}$.
After this block, $\bm{l}$ was concatenated to $\bm{h}$ and $\bm{c}$, and used as the initial hidden and cell state of the decoder LSTM.

%%%%%%%%%%
\vspace{3pt}
\noindent
{\bf Attention block:} Before calculating $p( w_n | \bm{x}, w_{1,..., n-1} )$ using the final FC layer, this block integrated the output of the LSTM $\bm{H} \in \mathbb{R}^{ (D+D_l) \times N}$ and $\bm{M}$ and outputs $\bm{M}' \in \mathbb{R}^{ (D+D_l) \times N}$ by using three FC-layers, like an MHSA with a single head.
Then the tanh activation was applied to $\bm{M}'$.
Finally, it is added to $\bm{H}$ as $\bm{H} + \tanh( \bm{M}' ) $ and passed to the final FC-layer to estimate $p( w_n | \bm{x}, w_{1,..., n-1} )$.

%%%%%%%%%%%%%%%%%%%%%%%%%%%%%%%%%%%%%%%%%%%%%%%%%%%%%%%%%%%
\subsection{Loss functions}

In order to train encoder/decoder and sub-blocks simultaneously, we designed loss function as a sum of multiple losses functions.
In addition, since the input audio and text were augmented by the mix-up, the cost function was also calculated using the mix-up; each loss was calculated for each of the two original label data and mixed using mixing-parameters $\beta$ and $(1 - \beta)$.
The following describes these loss functions in detail.

%%%%%%%%%%
\vspace{3pt}
\noindent
{\bf Word estimation loss:} For word prediction, we used the cross-entropy loss between $w_n$ and $p( w_n | \bm{x}, w_{1,..., n-1} )$. To avoid overfitting, we used label smoothing where smoothing factor was $0.1$.

%%%%%%%%%%
\vspace{3pt}
\noindent
{\bf Caption/meta keyword estimation loss:}
The weighted binary-cross entropy was used as the loss function for both caption/meta keyword estimation block as
\begin{align}
- \frac{1}{C^{\mbox{\tiny key}}} \sum_{i=1}^{C^{\mbox{\tiny key}}} 
\lambda_i z_i \ln p_i
+
\gamma_i (1-z_i) \ln (1 - p_i).
\end{align}
Note that for all variables, we omitted the superscripts ${}^{\mbox{\scriptsize cap}}$ and ${}^{\mbox{\scriptsize meta}}$ which indicate whether the variable belongs to caption keyword or meta keyword.
Here, the meanings of each variable are followings: 
$z_i$ is 1 when ground-truth keyword set includes $i$-th word and 0 otherwise, 
$p_i$ is the $i$-th value of the estimated posterior vector $\bm{p}$, and
$\lambda_i$ and $\gamma_i$ are the weight for $i$-th keyword as $\lambda_i = ( p(z_i) )^{-1}$ and $\gamma_i = ( 1 - p(z_i) )^{-1}$, respectively, where $p(z_i)$ is the prior probability of the $i$-th keyword calculated by
\begin{align}
p(z_i) = \frac{\mbox{\# of $c$-th keyword in training samples}}{ \mbox{\# of training samples} }.
\end{align}
To balance this loss and other losses, we multiplied a weight $(1-10^{-4})^{s}$ to this loss, where $s$ is the number of training steps.

%%%%%%%%%%
\vspace{3pt}
\noindent
{\bf Sentence length estimation loss:}
We used the softmax cross entropy between $L$ and $\bm{p}^{\mbox{\scriptsize len}}$ as the loss for the sentence length estimation block.
To balance this loss and other losses, we multiplied a weight $10^{-2}$ to this loss.

%%%%%%%%%%
\vspace{3pt}
\noindent
{\bf Keyword co-occurrence loss:}
In order to prevent the decoder outputs the words which are obviously not related to the meta keywords, we used the keyword co-occurrence loss between words in a caption and its meta keywords.
For example, when meta keywords are \{{\it car, sing, bird}\}, words not related to the keywords such as \{{\it people, children, talking, talk, speak}\} may not be included in the correct caption.
To prevent the decoder outputs such words, we adopted a penalty based on the decoder outputs $p( w_n | \bm{x}, w_{1,..., n-1} )$.

Before training, we created a hash-table of the co-occurrence lists; the keys of the hash-table are all keywords in the keyword vocabulary, and the element of each key is a list of the words that have co-occurred with the keyword in the training dataset.
For example, in the case of meta keywords are \{{\it car, sing, bird}\} and ground-truth captions are \{{\it Cars are driving and birds are singing, A car passes by while birds are chirping and singing}\}, 
\{{\it cars, are, driving, and, birds, singing, a, passes, by, while, chirping}\} are added to the co-occurrence lists of {\it car, sing}, and {\it bird}.
Then, in the training step, we added penalties of the decoder outputs to the whole loss value as
\begin{equation}
\frac{1}{C^{\mbox{\scriptsize cap}}}
\sum _{n=1}^N
\sum_{i=1}^{C^{\mbox{\scriptsize cap}}}
\lvert
b_i
\cdot
p( w_n = i | \bm{x}, w_{1,..., n-1} )
\rvert,
\end{equation}
where $b_i \in \{0, 1\} $ a binary mask where $b_i=1$ when none of all co-occurrence lists of the ground-truth meta keywords includes the $i$-th word, and otherwise $b_i=0$.

%%%%%%%%%%%%%%%%%%%%%%%%%%%%%%%%%%%%%%%%%%%%%%%%%%%%%%%%%%%
\begin{table*}[ttt]
\caption{Experimental results on development-testing dataset.}
\label{tab:result}
\centering
\begin{tabular}{ l | ccccccccc }
\toprule
\textbf{Model} 	& \textbf{B-1}	& \textbf{B-2}	& \textbf{B-3}	& \textbf{B-4}	& \textbf{CIDEr}	& \textbf{METEOR}	& \textbf{ROUGE-L} & \textbf{SPICE} & \textbf{SPIDEr} \\	
\midrule
$\mathtt{Baseline}$		& 38.9		& 13.6		& 5.5			& 1.5			& 7.4			& 8.4			& 26.2	& 3.3 	& 5.4 \\
\midrule 
$\mathtt{Model1}$		& 52.6		& 33.5		& 22.4			& 14.6      	& 30.1			& 14.7			& 34.7	& 9.0 	& 19.5 \\
$\mathtt{Model2}$		& 51.2		& 32.1		& 21.3			& 14.1			& 29.7			& 14.5			& 33.9	& 9.1 	& 19.4 \\
$\mathtt{Model3}$		& 51.7		& 33.0		& 22.0			& 14.5			& 30.0			& 14.7			& 34.3	& 8.6 	& 19.3 \\
$\mathtt{Model4}$		& 53.0		& 33.7		& 22.4			& 14.5			& 30.2			& 14.8			& 35.2	& 9.1 	& 19.6 \\
\midrule
$\mathtt{Ensemble}$		& $\bm{53.7}$&$\bm{34.8}$& $\bm{23.5}$   & $\bm{15.6}$	& $\bm{31.9}$	& $\bm{15.2}$	& $\bm{35.9}$	& $\bm{9.4}$	& $\bm{20.7}$	 \\
\bottomrule
\end{tabular}
\vspace{-5pt}
\end{table*}
%%%%%%%%%%%%%%%%%%%%%%%%%%%%%%%%%%%%%%%%%%%%%%%%%%%%%%%%%%%

%%%%%%%%%%%%%%%%%%%%%%%%%%%%%%%%%%%%%%%%%%%%%%%%%%%%%%%%%%%
\subsection{Beam search and test time augmentation}

We used the beam search decoding for the word decision process from $p( w_n | \bm{x}, w_{1,..., n-1} )$. The beam size was 5, and $n$-gram blocking size was 2, i.e. a hypothesis in a beam was discarded if there was a bi-gram that appeared more than once within it.
In addition, we used test time augmentation (TTA) for audio input.
This is because the audio input was randomly cropped for limiting the time-length as $T=216$ in training phase.
If the length of audio input is changed in testing phase, it may have a bad influence on the batch normalization layers.
Therefore, in testing-phase, we also randomly cropped and zero-padded the audio input so as to $T=216$.
We generated five input audios by this process, and took the average of five outputs of the decoder.

%%%%%%%%%%%%%%%%%%%%%%%%%%%%%%%%%%%%%%%%%%%%%%%%%%%%%%%%%%%
\subsection{Training hyper-parameters}
We used the AdamW \cite{adamw} optimier with a constant learning rate $10^{-4}$. The minibatch-size was 48.
We randomly splitted $2893 + 1045$ samples in the development dataset into 3842 training samples and 96 validation samples.
We used a DNN whose validation score was the best while 300 epochs training.

%%%%%%%%%%%%%%%%%%%%%%%%%%%%%%%%%%%%%%%%%%%%%%%%%%%%%%%%%%%
\subsection{Submitted systems}

We used a model ensemble to output the final results; each model in the ensemble outputted $\ln p( w_n | \bm{x}, w_{1,..., n-1} )$, and we took the average of all log-probabilities in the beamsearch phase.
The four submitted results were four types of different combinations of following models.
\begin{description}
 \setlength{\parskip}{0cm} 
 \setlength{\itemsep}{0.1cm} 
% model 11
\item[$\mathtt{Model1}$] The base model described in Sec. 2.3.
% model 10
\item[$\mathtt{Model2}$] Modified model of $\mathtt{Model1}$. A FC layer was used instead of the MHSA layer in the caption keyword estimation block.
% model 9
\item[$\mathtt{Model3}$] Modified model of $\mathtt{Model2}$. The meta keyword estimation block in the encoder and the attention block in the decoder were removed.
% model 8
\item[$\mathtt{Model4}$] Modified model of $\mathtt{Model1}$. Mix-up augmentation for text was removed.
% model 12
\item[$\mathtt{Model5}$] Modified model of $\mathtt{Model1}$. The audio embedding block consists of 
one CNN block, the reshape block, one FC layer for changing the hidden dimension to $D=120$, and one shared Transformer encoder block \cite{transformer} with time-direction sub-sampling operation. The Transformer encoder block and sub-sampling operation were used twice, with the sub-sampling operation thinning out the one time-frame every two time-frames. 
% model 13
\item[$\mathtt{Model6}$] Modified model of $\mathtt{Model1}$. The encoder has only one BLSTM layer, and $D=160$.
\end{description}
The details of four submitted systems are followings:
\begin{description}
 \setlength{\parskip}{0cm} 
 \setlength{\itemsep}{0.1cm} 
 % ensemble #4
\item[Submission 1] Ensemble of 20 models. This model consists of
two $\mathtt{Model1}$,
two $\mathtt{Model1single}$,
two $\mathtt{Model1param2}$,
two $\mathtt{Model2}$,
two $\mathtt{Model3}$,
two $\mathtt{Model3param2}$,
two $\mathtt{Model4}$, 
two $\mathtt{Model4single}$, and 
four $\mathtt{Model4param2}$.
The number of trainable parameters was 33.0M.
 % ensemble #6
\item[Submission 2] Ensemble of 50 models. This model consists of
five $\mathtt{Model1}$,
five $\mathtt{Model1single}$,
five $\mathtt{Model1param2}$,
five $\mathtt{Model2}$,
five $\mathtt{Model3}$,
five $\mathtt{Model3param2}$,
five $\mathtt{Model4}$, 
five $\mathtt{Model4single}$, and 
ten $\mathtt{Model4param2}$.
The number of trainable parameters was 82.5M.
 % ensemble #5
\item[Submission 3] Ensemble of 12 models. This model consists of
two $\mathtt{Model1}$,
two $\mathtt{Model3}$,
four $\mathtt{Model4}$, 
two $\mathtt{Model5single}$, and 
two $\mathtt{Model6single}$.
The number of trainable parameters was 20.7M.
 % ensemble #7
\item[Submission 4] Ensemble of 30 models. This model consists of
five $\mathtt{Model1}$,
five $\mathtt{Model3}$,
ten $\mathtt{Model4}$, 
five $\mathtt{Model5single}$, and 
five $\mathtt{Model6single}$.
The number of trainable parameters was 51.7M.
\end{description}
where $\mathtt{single}$ means we did not used the HPSS (i.e. $\bm{X} =\bm{S}$),
and $\mathtt{param2}$ means two additional modification: 
(i) before adding the meta keyword estimation loss, we multiplied 0.8 to it as a loss weight.
(ii) we did not used the second trick in minibatch sample selection, i.e. the target caption was selected with equal probability from the five ground-truth caption of an audio.

\section{Evaluation on dev-test dataset}

To give a sense of the accuracy of the submitted system, we tested a simplified {\bf Submission 1} on the development-test dataset of the Challenge.
First, we conducted three unit tests for $\mathtt{Model1}$, $\mathtt{Model2}$, $\mathtt{Model3}$, and $\mathtt{Model4}$, and then evaluated the ensemble model as $\mathtt{Ensemble}$.
Although $\mathtt{Ensemble}$ is simpler than actual our challenge submissions, it should be useful for testing the performance of each model and the effectiveness of the ensemble.

Table \ref{tab:result} shows the evaluation results.
All models significantly outperformed the baseline system, and with these ensembles model achieved the SPIDEr score $20.7$.
Our model consists of a complex combination of various sub-blocks and cost functions.
As a future work, we will conduct ablation studies to determine how each blocks/cost functions has affected.

\section{Conclusions}

This technical report described the system participating to the DCASE 2020 Challenge Task 6 \cite{task}. Our submission focused on solving the indeterminacy problems in word selection and sentence length. We simultaneously solved the main caption generation and sub indeterminacy problems by estimating keywords and sentence length through multi-task learning. The SPIDEr score of our submission on the development-testing dataset was $20.7$. Since our model consisted of a complex combination of various sub-blocks and cost functions, as a future work, we will conduct ablation studies for these modules.

% -------------------------------------------------------------------------
% Either list references using the bibliography style file IEEEtran.bst

\end{sloppy}
\end{document}